\documentclass[pra,twocolumn,showpacs,superscriptaddress]{revtex4-1}
\usepackage{graphicx}
\usepackage{amssymb}
\usepackage{amsfonts}
\usepackage{amsmath}
\usepackage{lmodern}
\usepackage{mathrsfs}
\usepackage[dvipdfm,colorlinks=true, citecolor=blue, urlcolor=blue ]{hyperref}
\begin{document}
\title{Resonance fluorescence beyond the dipole approximation of a quantum dot in a plasmonic nanostructure}
\author{Chun-Jie Yang}
\affiliation{Center for Interdisciplinary Studies and Key Laboratory for Magnetism and Magnetic Materials of the MoE, Lanzhou University, Lanzhou 730000, China}
\author{Jun-Hong An}\email{anjhong@lzu.edu.cn}
\affiliation{Center for Interdisciplinary Studies and Key Laboratory for Magnetism and Magnetic Materials of the MoE, Lanzhou University, Lanzhou 730000, China}
\begin{abstract}
The mesoscopic characteristics of a quantum dot (QD), which make the dipole approximation (DA) break down, provide a new dimension to manipulate light-matter interaction [M. L. Andersen, \textit{et al.}, Nat. Phys. \textbf{7}, 215 (2011)]. Here we investigate the power spectrum and the second-order correlation property of the fluorescence from a resonantly driven QD placed on a planar metal. It is revealed that due to the pronounced QD spatial extension and the dramatic variation of the triggered surface plasmon near the metal, the fluorescence has a notable contribution from the quadrupole moment. The $\pi$-rotation symmetry of the fluorescence to the QD orientation under the DA is broken. By manipulating the QD orientation and quadrupole moment, the spectrum can be switched between the Mollow triplet and a single peak, and the fluorescence characterized by the antibunching in the second-order correlation function can be changed from the weak to the strong radiation regime. Our result is instructive for utilizing the unique mesoscopic effects to develop nanophotonic devices.
\end{abstract}
\maketitle
\section{introduction}
Quantum optics has advanced to the stage of experimental measurement and manipulation of individual quantum systems in single quanta level \cite{Haroche2013a,Wineland2013,Koelemeij2014,Bloom2014,Wang2015,Kato2015}, where the light-matter interaction plays an important role. Considerable interest has been generated in exploring new mechanisms that enable efficient control of the light-matter interaction. In past years, with the sufficient reduction of the effective mode volume for photons, strong and even ultrastrong light-matter interaction have been experimentally realized \cite{Wallraff2004,Guebrou2012,Niemczyk2010,Scalari2012,Tudela2013,Cacciola2014,Torma2015}. Recently, a scheme exploiting the mesoscopic characteristics of quantum dots (QDs) was proposed \cite{Andersen2011}, by means of which the plasmon-matter interaction \cite{Akimov2007} can be strongly modified.

The light-matter interaction is generally described under the dipole approximation (DA), which works well in atomic systems where the variation of the field is negligible within the atomic spatial extension \cite{Cronin2009,Haroche2013b,Muller2007,Monniello2013}. However, once the spatial variation of the field becomes pronounced, such as the surface plasmons triggered by the radiation field of the quantum emitter \cite{Zayatsa2005}, and the emitter is spatially extended, such as a QD several tens of nanometers in size \cite{Yoshie2004,Lodahl2015}, the validity of the DA is not clear \textit{a priori}. Experimentally, a large deviation from the dipole theory was observed for QDs in close proximity to a silver mirror \cite{Andersen2011}. The optical response of quantum nanosystems beyond the long-wavelength approximation (equivalent to the DA) has been studied semiclassically \cite{Ajiki2002,Cho2003,Bamba2008,Iida2009}, which indicated that the nonlocal spatial interplay between the wave functions of the QD exciton and the electromagnetic field makes the DA invalid. Unconventional phenomena exceeding the DA have been found, such as the selection-rule breakdown of an isolated single-walled carbon nanotube in a nanogap \cite{Iida2011,Takase2013}, entangled-photon generation from biexcitons in a semiconductor film \cite{Bamba2010}, and enhanced up-conversion of entangled photons in nanostructures \cite{Osaka2012}. In the fully quantum theory, making Taylor expansion of the field spatial distribution function to the first order, it is found that the nonlocal interaction is described by the quadrupole moment, and a microscopic picture of it from a circular quantum current density flowing along a curved path inside the QD has been provided \cite{Stobbe2012,Tighineanu2015}. Furthermore, as the quadrupole moment can be tuned by controlling the size and shape of the QD \cite{Johansen2008,Leistikow2009}, it has potential applications in the development of a nanophotonic devices. For the requirement of developing nanoplasmonic single-photon source, a study on the fluorescence from a resonantly driven QD, especially the second-order correlation property of the fluorescence, modified by the mesoscopic characteristics is necessary and important.

In this work, we study the resonance fluorescence of a mesoscopic QD in different spatial orientations placed near a plasmonic nanostructure. Going beyond the DA, a microscopic description to the power spectrum and the second-order correlation property of the QD fluorescence is established. The substantial deviations from the dipole theory are found when the QD is positioned within the penetration depth of the plasmons into the dielectric. It is revealed that the spatial rotation symmetry for the resonance fluorescence spectrum over the QD orientation is changed from $\pi$ under the DA to $2\pi$ due to the interference of the emission from the dipole and the quadrupole moments. The widths and intensities of the spectral peaks differ dramatically from those under the DA due to the cooperative actions of the dipole and quadrupole moments. Explicitly, by exploiting the QD mesoscopic effects, the spectrum can be switched between a single peak and the Mollow triplet. The analysis on the second-order correlation property indicates that, keeping the nonclassical antibunching nature, the fluorescence can be changed from the weak to the strong emission regime by increasing the quadrupole moment. This opens an avenue to develop nanophotonic single-photon devices by use of the QD mesoscopic characters. Our parameter values are all experimentally attainable.

Our paper is organized as follows. In Sec. \ref{model}, we show the model and establish a microscopic description to the QD--surface-plasmon interaction in arbitrary QD orientations beyond the DA. In Sec. \ref{fluorescence}, the fluorescence spectrum and the second-order correlation property are numerically studied. In Sec. \ref{con}, a summary is given.
\section{QD--surface-plasmon interaction beyond the DA}
\label{model}
\subsection{System and Green's tensor}
\begin{figure}[tbp]
\includegraphics[width=0.9\columnwidth]{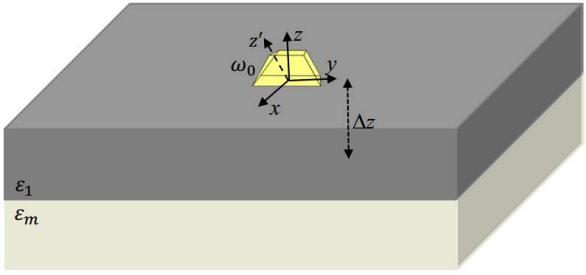}
\caption{Diagram of a QD with characteristic frequency $\omega_0$ embedded in gallium arsenide at a distance $\Delta z$ above a metal. $\varepsilon_1$ and $\varepsilon_m$ are dielectric constants of the media. $z$ and $z^\prime$ represent different orientations of the QD.} \label{Fig1}
\end{figure}
Our system is depicted in Fig. \ref{Fig1}: a QD with frequency $\omega_0$ embedded in gallium arsenide (GaAs) media is placed on a dissipative metal. The metal is characterized by a complex Drude dielectric function $\varepsilon_m (\omega )=\varepsilon _{\infty }[1-\frac{\omega _{p}^{2}}{\omega (\omega +i\gamma _{p})}]$, where $\omega _{p}$ is the bulk plasma frequency, $\varepsilon _{\infty}$ is the high-frequency limit of the metal dielectric function, and $\gamma _{p}$ represents the Ohmic loss responsible for the dissipation of the electromagnetic field in the metal. Here the metal is chosen as silver with the parameters $\omega_{p}=3.76$ eV, $\varepsilon_{\infty}=9.6$, $\gamma_{p}=0.03\omega_{p}$ in the interested frequency range, and the dielectric permittivity of GaAs is $\varepsilon_1=12.25$ \cite{Johnson1972,Tudela2010}. We assume that the layered media are linear, isotropic and nonmagnetic ($\mu_{1}=\mu_{\text{m}}=1$).

The electromagnetic field in dispersive and absorbing dielectrics is described by the Green's tensor $\mathbf{G}(\mathbf{r},\mathbf{r}^{\prime };\omega )$ \cite{Dung1998,Sondergaard2001}, which is rendered as the field in frequency $\omega$ evaluated at $\mathbf{r}$ due to a point source at $\mathbf{r}^{\prime}$. It can be obtained by solving the Maxwell-Helmholtz wave equation $[{\pmb\nabla}\times{\pmb\nabla}\times-\frac{\omega^{2}}{c^{2}}\varepsilon(\omega)]\mathbf{G}(\mathbf{r},\mathbf{r}^{\prime };\omega )=\mathbf{I}\delta (\mathbf{r}-\mathbf{r}^{\prime })$, where $\mathbf{I}$ is identity matrix. For general geometries, the solving needs some numerical methods, such as the finite difference time domain and the finite element methods \cite{Chen2010b,Cano2010}. For symmetric geometries such as spheres, cylinders, or planes, its analytical solution is achievable \cite{Novotny2006,Dzsotjan2010,Zubairy2014}. In our configuration, $\mathbf{G}(\mathbf{r},\mathbf{r}^{\prime };\omega )$ in the upper half-space of the metal-dielectric interface is calculated as the sum of the free-space and reflected Green's tensors $\mathbf{G}(\mathbf{r},\mathbf{r}^{\prime },\omega )=\mathbf{G}_{0}(\mathbf{r},\mathbf{r}^{\prime },\omega )+\mathbf{G}_{\text{R}}(\mathbf{r},\mathbf{r}^{\prime},\omega )$. See more details in Appendix \ref{Green}.

Three distinct modes are triggered by the emission of the QD. The first one is the radiative modes propagating into the free space. The second one is the damped non-radiative mode due to the Ohmic loss in the metal. The last one is the tightly confined field called surface plasmon propagating along the metal surface \cite{Pitarke2007}. The electromagnetic modes associated with the surface plasmon enable strong confinement of light on the surface and thus enhance the light-matter interaction, which has inspired great interests in studying surface plasmon subwavelength optics and quantum plasmonics \cite{Barnes2003,Garcia2010,Tame2013}. In addition, due to the exponential decay of the intensity of the electromagnetic field perpendicular to the metal surface, its variation along this direction is pronounced within the spatial extension of the QDs. It causes the breakdown of the DA. Thus a new theory in describing light-matter interaction beyond the DA is necessary.

\subsection{QD--surface-plasmon interaction beyond the DA}\label{int-b-da}
The QD-field interaction is described by the minimal coupling Hamiltonian $\hat{H}_{\text{int}}(\mathbf{r},t)=-\frac{q}{m}\mathbf{A}(\mathbf{r},t)\cdot \mathbf{\hat{p}}$, where $\mathbf{\hat{p}}$ is the momentum operator, $q$ and $m$ are the electronic charge and mass, respectively, and $\mathbf{A}(\mathbf{r},t)$ is the vector potential of the field \cite{Stobbe2012}. In quantization, $\mathbf{A}(\mathbf{r},t)$ is expanded as $\mathbf{A}(\mathbf{r},t)=\sum_{l}\sqrt{\frac{\hbar }{2\omega_{l}\varepsilon _{0}}}[\mathbf{A}_{l}(\mathbf{r})\hat{a}_{l}e^{-i\omega_{l}t}+\text{h.c.}]$, where $\mathbf{A}_l(\mathbf{r})$ relevant to the Green's tensor is the field spatial distribution function, $\hat{a}_{l}$ is the annihilation operator with frequency $\omega_{l}$, $\varepsilon_0$ is the vacuum dielectric function, and $l=(\mathbf{k},s)$ is the combined index of the wave vector $\mathbf{k}$ and polarization $s\in(1,2)$. To go beyond the DA, we make a Taylor expansion of $\mathbf{A}_{l}\mathbf{(r)}$ to the first order around the QD center
\begin{equation}
\mathbf{A}_{l}(\mathbf{r})\simeq \mathbf{A}_{l}(\mathbf{r}_{0})+(\mathbf{r}-\mathbf{r}_{0})\cdot \pmb{\mathfrak{J}}\mathbf{A}_{l}(\mathbf{r})|_{\mathbf{r}=\mathbf{r}_{0}},
\end{equation}
where $\pmb{\mathfrak{J}}\mathbf{A}_{l}(\mathbf{r})$ is the Jacobian matrix of partial derivatives of $\mathbf{A}_{l}(\mathbf{r})$. The QD in the strong confinement regime can be well described by a two-band model with states $| c\rangle$ and $| v\rangle$ representing an electron and a hole in the conduction and heavy valence band, respectively \cite{Stobbe2009}. Employing the rotating wave approximation, we arrive at the interaction Hamiltonian beyond the DA in the interaction picture
\begin{equation}
\hat{H}_{\text{I}}(t)=\hbar \sum_{l}(g_{l}e^{i\Delta _{l}t}\hat{\sigma}_{-}\hat{a}_{l}^{\dag }+\text{H.c.}), \label{H-inter0}
\end{equation}
where $\hat{\sigma}_{-}=|v\rangle \langle c |$, and $\Delta_l=\omega_{l}-\omega_{0}$ is the frequency detuning. The QD-field coupling strength is
\begin{equation}
g_{l}=-\frac{q}{m}\sum_{j,k}(\frac{1}{2\hbar \epsilon _{0}\omega _{l}})^{1/2}[(\mu _{j}+\Lambda _{j,k}\nabla _{k})A_{lj}^{\ast }(\mathbf{r})]_{\mathbf{r=r}_{0}}, \label{couple}
\end{equation}
where $j$ and $k$ index the three Cartesian coordinates $x,y,z$, $\mu_j=\langle v|\hat{p}_j|c\rangle $ and $\Lambda_{j,k}=\langle v|\hat{p}_{j}r_k|c\rangle $ denote the dipole and the quadrupole moments of the QD, respectively, and $\nabla _{k}$ represents the differential of $A_{lj}^{\ast }(\mathbf{r})$ to the $k$th coordinate component.

One can see from Eq. (\ref{couple}) that both the dipole and quadrupole moments contribute to its interaction with the radiation field. The former couples to the field distribution function $\mathbf{A}_l(\mathbf{r})$, while the latter couples to the gradient of $\mathbf{A}_l(\mathbf{r})$. In atomic systems, the atom is much smaller than the wavelength and the typical length of the spatial distribution of the field, i.e., $\nabla_k A_{l,j}^{\ast }(\mathbf{r})|_{\mathbf{r}=\mathbf{r}_0} \simeq 0$. Thus the contributions from the quadrupole moment can be safely abandoned and the DA is applicable. However, in the QD system, as the QD is large in size and the spatial variation of the field is pronounced, they cannot be ignored and the DA is inapplicable.

The interaction between the QD and the field is further characterized by the spectral density $J(\omega )=\sum_{l}g_{l}^{2}\delta (\omega -\omega _{l})$. Combined with Eq. (\ref{couple}), it takes the form
\begin{align} \label{spectral}
\begin{split}
J (\omega) &=\frac{q^{2}}{\pi c^{2}\hbar ^{2}\varepsilon_{0}m^{2}}\sum_{j,n,j^{\prime },n^{\prime }}\{(\mu _{j}+\Lambda_{j,n}\nabla _{n})\\
&\times (\mu _{j^{\prime }}^{\ast }+\Lambda _{j^{\prime },n^{\prime }}^{\ast }\nabla_{n^{\prime }}^{\prime })\text{Im}[G_{j,j^{\prime }}(\mathbf{r},\mathbf{r}^{\prime };\omega )]\}_{\mathbf{r=r}^{\prime }=\mathbf{r}_{0}} ,
\end{split}
\end{align}
where $G_{j,j^{\prime }}(\mathbf{r},\mathbf{r}^{\prime };\omega )$ is the $(j,j^\prime)$ element of the Green's tensor and the relation $\textrm{Im}[G_{j,j^{\prime }}(\mathbf{r,r}^{\prime };\omega )]=\frac{\pi c^{2}}{2\omega }\sum_{l}A_{l,j}^{\ast }(\mathbf{r})A_{l,j^{\prime }}(\mathbf{r}^{\prime })\delta (\omega -\omega _{l})$ has been utilized \cite{Stobbe2012}. In the past decade, a division of the Green's tensor into the surface plasmons bounded on the surface and the out-of-plane waves propagating away from the surface has been studied \cite{Sondergaard2004,Siahpoush2012,Siahpoush2014}. In the following, we shall study the spectral density for different orientations of the QD.
\subsection{Spectral density in arbitrary QD orientations}
Consider first the special case that the QD orientates in $z$ axis. According to the symmetry of the electron and hole wavefunctions, we can calculate the two moments $\vec{\mu}=\bar{\mu}\left(\begin{array}{ccc}1 &i &0\end{array}\right)^T$ and $\mathbf{\Lambda }=\bar{\Lambda}\left(\begin{array}{ccc}0 & 0 & 0 \\0 & 0 & 0 \\1 & i & 0\end{array}\right)$, where $\bar{\mu}$ and $\bar{\Lambda}$ can be fitted experimentally \cite{Andersen2011}. When the QD orientates in $z^\prime$ shown in Fig. \ref{Fig1}, which can be expressed as a $\phi$-rotation of the QD along the $x$ axis from the $z$ direction, it can be proved that the QD-field interaction Hamiltonian (\ref{H-inter0}) is unchanged except that the moments change into $\vec{\tilde{\mu}}(\phi)=\langle c|\hat{U}_{x}^{\dag }(\phi)\mathbf{\hat{p}}\hat{U}_{x}(\phi )|v\rangle $ and $\tilde{\mathbf{\Lambda}}(\phi)=\langle c|\hat{U}_{x}^{\dag }(\phi)\mathbf{\hat{p}r}\hat{U}_{x}(\phi )|v\rangle $, where $\hat{U}_{x}(\phi )=e^{-(i/\hbar)\hat{L}_{x}\phi}$ with $\hat{L}_{x}$ the QD angular momentum and $\phi$ the angle between $z$ and $z^\prime$. We thus have the moments
\begin{eqnarray}
\vec{\tilde{\mu}}(\phi) &=&\bar{\mu}\left(
\begin{array}{ccc}
1 &
i\cos \phi  &
i\sin \phi
\end{array}%
\right)^T , \label{m1}\\
\tilde{\mathbf{\Lambda}}(\phi) &=&\bar{\Lambda}\left(
\begin{array}{ccc}
0 & 0 & 0 \\
-\sin \phi  & -i\sin \phi \cos \phi  & -i\sin \phi \sin \phi  \\
\cos \phi  & i\cos \phi \cos \phi  & i\cos \phi \sin \phi
\end{array}%
\right) .\label{m2}
\end{eqnarray}
Inserting Eqs. (\ref{m1}) and (\ref{m2}) into Eq. (\ref{spectral}), we obtain the spectral density
\begin{equation}
J(\omega)=J _{0}(\omega)+J _{\text{R}}(\omega),
\end{equation}
where $J _{0}(\omega)$ and $J _\text{R}(\omega)$ contain the contributions from the free-space field $\mathbf{G}_{0}(\mathbf{r},\mathbf{r}^{\prime },\omega )$ and the reflected field $\mathbf{G}_\text{R}(\mathbf{r},\mathbf{r}^{\prime },\omega )$, respectively. Their forms in the cylindrical coordinate are
\begin{eqnarray}
&J_{0}(\omega) =\frac{\omega }{\Phi }\int ds\textrm{Re}[A_{1}\bar{\mu}^{2}+B_{1}\bar{\Lambda}^{\prime 2}],& \\
&J_{\text{R}}(\omega ) =\frac{\omega }{\Phi }\int ds\textrm{Re}\{[A_{2}\bar{\mu}^{2}+B_{2}\bar{\Lambda}^{\prime 2}+B_{3}\bar{\mu}\bar{\Lambda}^{\prime }]e^{2ik_{z_1}\Delta z}\}.~~~&\label{ref}
\end{eqnarray}
where $\Phi =8\pi^2 \varepsilon _{0}m^{2}\hbar^{2}c^{3}/(q^{2}n_{1})$, $\bar{\Lambda}^{\prime}=k_{1}\bar{\Lambda}$ with $k_1=n_1\omega/c$, $s=k_{\rho}/k_{1}$, and $s_{z}\equiv\sqrt{1-s^{2}}=k_{z_{1}}/k_{1}$. The coefficients are given as
\begin{eqnarray}
A_{1} &=&\frac{s}{s_{z}}[(2-s^{2})(1+\cos ^{2}\phi )+2s^{2}\sin^{2}\phi ], \label{aa}\\
A_{2}&=&\frac{s}{s_{z}}[(r^{\text{s}}-r^{\text{p}}s_{z}^{2})(1+\cos ^{2}\phi)+2 r^{\text{p}}s^{2}\sin ^{2}\phi], \\
B_{1} &=&\frac{s}{s_{z}}\{(2-10s^2+{35s^4\over 4})\sin ^{4}\phi+2 s^2(4-5s^2)\nonumber\\
&&\times\sin^{2}\phi+2s^4\}, \label{bb}\\
B_{2}&=&\frac{s}{s_{z}}\{[r^\text{p}-r^\text{s}+\frac{3}{4}(r^\text{p}+r^\text{s}-r^\text{p}s^2)s^2]\sin^4\phi\nonumber\\
&&+s^2(r^\text{s}-3r^\text{p})\sin^2\phi+2r^\text{p}s^4 \},\\
B_{3} &=&2is[2 r^{\text{p}}s^{2}+(r^{\text{p}}s^{2}+r^{\text{p}}-r^{\text{s}})\sin ^{2}\phi]\cos \phi,
\end{eqnarray}
where $r^{\text{s}}=\frac{s_{z}-\sqrt{n_{m1}^{2}-s^{2}}}{s_{z}+\sqrt{n_{m1}^{2}-s^{2}}}$ and $r^{\text{p}}=\frac{\varepsilon (\omega )s_{z}-\varepsilon _{1}\sqrt{n_{m1}^{2}-s^{2}}}{\varepsilon (\omega )s_{z}-\varepsilon _{1}\sqrt{n_{m1}^{2}-s^{2}}}$ are the Fresnel reflection coefficients for s- and p-polarized lights with the relative dielectric function $n_{m1}=\sqrt{\varepsilon_m(\omega)/\varepsilon_1}$.

Up to now, going beyond the DA, we have analytically established the microscopic description to the interaction between the QD and the radiation field propagating near a plasmonic nanostructure. It can be seen that $J_{\text{R}}(\omega )$ is contributed from two types of field modes along the $z$-direction by dividing the integration range $[0,\infty]$ of $s$ in Eq. (\ref{ref}) into two intervals $[0,1]$ and $[1,\infty]$. The former has a real $k_{z_1}$ and is associated with the reflected plane waves by the metal-dielectric interface, while the latter has a complex $k_{z_1}$ and is associated with the surface plasmons and the damped non-radiative mode \cite{Chang2006,Tudela2014}. Furthermore, the cooperative effect of the dipole and the quadrupole moments are self-consistently contained in $J_{\text{R}}(\omega )$, where the $B_3$ term characterizes the interference between the two moments. Just due to this interference, the decoherence of the QD shows significant differences from the one under the DA. Under the DA, the spectral density contributed uniquely from the dipole moment has a $\pi$-rotation symmetry over the QD orientation. When the quadrupole moment is taken into account, the symmetry is changed into $2\pi$ because of the presence of the $B_{3}$ term. It is noted that in the special case $\phi=0$ or $\pi$, our result reduces exactly to the one in Ref. \cite{Andersen2011}.

\section{Fluorescence modified by the QD mesoscopic effects} \label{fluorescence}
We consider explicitly that the QD is resonantly driven by a laser so that the resonance fluorescence of the QD near the metal surface beyond the DA can be measured. In a frame rotating at the laser frequency $\omega_0$, the master equation under the Born-Markovian approximation reads
\begin{align}\begin{split}\label{master}
\dot{\rho}(t)=&-i\frac{\Omega}{2} [\hat{\sigma}_{+}+\hat{\sigma}_{-},\rho (t)]\\
&+\frac{\Gamma}{2}[2\hat{\sigma}_{-}\rho (t)\hat{\sigma}_{+}-\hat{%
\sigma}_{+}\hat{\sigma}_{-}\rho (t)-\rho (t)\hat{\sigma}_{+}\hat{\sigma}_{-}],
\end{split}\end{align}
where $\Omega$ is the Rabi frequency denoting the laser pumping strength and $\Gamma =2\pi J(\omega_0 )$ is the QD spontaneous emission rate.

\begin{figure}[tbp]
\includegraphics[width=\columnwidth]{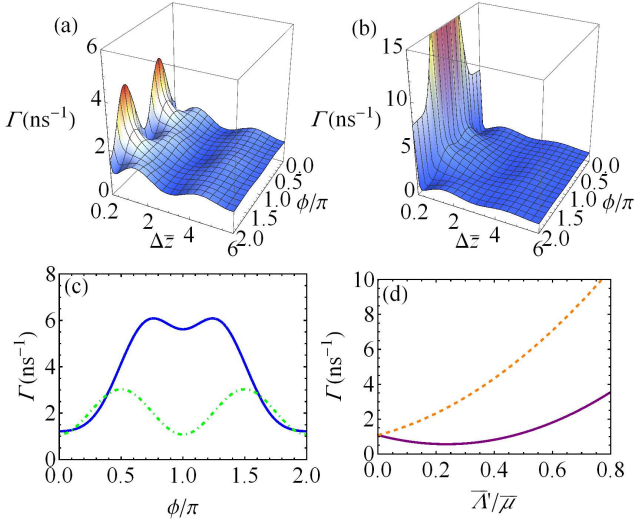}
\caption{Orientation dependence of $\Gamma$ for different dimensionless separation $\Delta \bar{z}$ under (a) and beyond (b) the DA. (c) A cross section of (a) and (b) at $\Delta \bar{z}=0.3$ [blue solid and green dot-dashed lines from (a) and (b), respectively]. (d): $\Gamma$ as a function of $\bar {\Lambda}^\prime/\bar{\mu}$ when $\phi=0$ (purple solid line) and $\phi=\pi$ (orange dashed line). Parameters are $\omega_0=1.2$ eV, $\Phi=3.0\times10^{9}\bar{\mu}^2$, and $\bar{\Lambda}^{\prime }/\bar{\mu}=0.5$, which are obtained by fitting the experimental result in Ref. \cite{Andersen2011}.} \label{Fig2}
\end{figure}

In Figs. \ref{Fig2}(a) and \ref{Fig2}(b) we plot $\Gamma$ in different QD orientations as a function of the QD-interface separation $\Delta \bar{z}=\Delta z\omega _{p}/c$ under and beyond the DA, respectively. It can be seen that $\Gamma$ attenuates rapidly at small $\Delta\bar{z}$ and tends to a persistent oscillation with the increase of $\Delta\bar{z}$. This is understandable based on Eq. (\ref{ref}), which reveals that the contribution from the surface plasmons only dominates the small $\Delta \bar{z}$ regime. Beyond this regime, the spectral density originates mainly from reflected plane waves, which shows a lossless oscillation with the increase of $\Delta \bar{z}$. In addition, the significant deviations to the result under DA can be seen at small $\Delta \bar{z}$, where the surface plasmons bounded around the metal surface play a significant role. The typical distance where this deviation is observable is the penetration depth of the plasmons into the dielectric, which takes $\Delta \bar{z}_\text{c}\thicksim2$ for our parameters. Therefore, the DA is inapplicable especially when the QD is positioned within $\Delta \bar{z}_\text{c}$. This has been verified experimentally \cite{Andersen2011}. A cross section view at $\Delta \bar{z}=0.3$ is plotted in Fig. \ref{Fig2}(c). It indicates clearly that $\Gamma$ experiences a $\pi$ rotation symmetry over the QD orientation under the DA, while it is changed to $2\pi$ once the mesoscopic effects are taken into account. This agrees with our analytical expectation. To evaluate explicitly the  mesoscopic effects, we plot in Fig. \ref{Fig2}(d) $\Gamma$ as a function of $\bar{\Lambda}^\prime/\bar{\mu}$ at $\phi=0$ and $\pi$. It indicates that the interference between the dipole and quadrupole moments can cause a constructive increase or a destructive decrease of the decay rate under the DA (i.e., the value when $\bar{\Lambda}^\prime=0$). It demonstrates that we can control the QD decay by manipulating the mesoscopic characteristics of the QD, such as its spatial orientation and quadrupole moment.

The resonance fluorescence spectrum of the driven QD is defined as $S(\omega )=\frac{I_0}{\pi }\textrm{Re}[\int_{0}^{\infty }d\tau e^{i\omega \tau
}\langle \hat{\sigma}_{+}(t) \hat{\sigma}_{-}(t+\tau )\rangle _{\text{ss}}]$, where ``ss" denotes the steady state and $I_0(\mathbf{r})$ depending on the distance between the detector and the QD is a constant. From the master equation (\ref{master}) and with the use of the quantum regression theorem, the spectrum admits an analytical expression \cite{Carmichael2000}. It consists of the coherent (Rayleigh scattering) and incoherent (inelastic scattering) components. The coherent one is a delta function and ignored here, while the incoherent one takes the form
\begin{align}\begin{split}\label{spectrum}
&S(\delta\omega ) =\frac{Y^{2}}{8(1+Y^{2})}\Big[\frac{\Gamma }{\delta \omega^{2}+\frac{\Gamma ^{2}}{4}}\\
&~~~+\frac{\frac{3\Gamma }{4}P-(\delta \omega -\alpha )Q}{(\delta\omega -\alpha)^{2}+(\frac{3\Gamma }{4})^{2}}+\frac{\frac{3\Gamma }{4}P+(\delta\omega +\alpha )Q}{(\delta\omega +\alpha)^{2}+(\frac{3\Gamma }{4})^{2}}\Big],
\end{split}\end{align}
where $\delta\omega=\omega -\omega_{0}$, $Y=\frac{\sqrt{2}\Omega }{\Gamma }$, $i\alpha =\frac{\Gamma }{4}\sqrt{1-8Y^{2}}$, $P=\frac{Y^{2}-1}{Y^{2}+1}$, and $Q=\frac{\Gamma }{4\delta }\frac{1-5Y^{2}}{1+Y^{2}}$.
In the strong-driving and weak-radiation situation ($\Omega>\Gamma/4$), the spectrum constitutes of a sum of three Lorentzian components centered at $\omega_0$ and $\omega_0\pm \Omega$, respectively. This is the typical feature of the Mollow's triplet structure. In the weak-driving and strong-radiation situation ($\Omega<\Gamma/4$), the two sideband peaks disappear. Furthermore, as shown in Eq. (\ref{spectrum}), the positions, the widths, and the intensities of the three peaks of the spectrum are all associated with the decay rate $\Gamma$. Therefore, the spectrum can be greatly influenced by the mesoscopic effects of the QD via $\Gamma$.

\begin{figure}[tbp]
\includegraphics[width=\columnwidth]{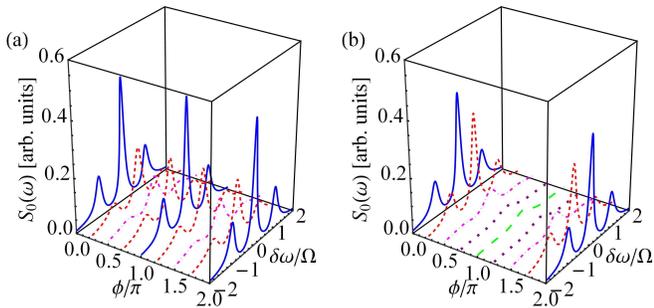}
\caption{Incoherent spectrum $S_0(\omega)=S(\omega)/I_0$ in different QD orientations under (a) and beyond (b) the DA. The parameters are the same as in Fig. \ref{Fig2}(c) except for $\Omega=5$ ns$^{-1}$.} \label{Fig3}
\end{figure}
\begin{figure}[tbp]
\includegraphics[width=\columnwidth]{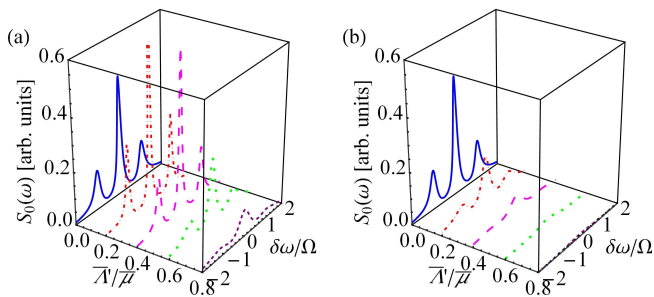}
\caption{Incoherent spectrum $S_0(\omega)=S(\omega)/I_0$ in different $\bar{\Lambda}^\prime/\bar{\mu}$ for $\phi=0$ (a) and $\pi$ (b). The parameters are the same as in Fig. \ref{Fig2}(d) except for $\Omega=5$ ns$^{-1}$.} \label{Fig4}
\end{figure}
Figures \ref{Fig3}(a) and \ref{Fig3}(b) show the spectrum in different QD orientations under and beyond the DA, respectively. The spectrum under the DA has a $\pi$ rotation symmetry over the QD orientation, while beyond the DA, it has $2\pi$ symmetry. It agrees with the behavior of $\Gamma$, Fig. \ref{Fig2}(c). Another interesting observation is that although the spectrum keeps the Mollow triplet structure in the whole range of $\phi$ when the DA is applied [see Fig. \ref{Fig3}(a)], it can be switched from the Mollow triplet to a single peak centered at $\delta\omega=0$ by adjusting the QD orientation when the mesoscopic effect is considered [see Fig. \ref{Fig3}(b)]. This result manifests clearly the anisotropy of the spontaneous emission of the QD placed on the metal surface. Figure \ref{Fig4} plots the spectrum with the change of $\bar{\Lambda}^\prime/\bar{\mu}$ for $\phi=0$ (a) and $\pi$ (b), respectively. When $\phi=0$, the Mollow triplet can be either strengthened or weakened with the increasing of $\bar{\Lambda}^\prime/\bar{\mu}$, while at $\phi=\pi$, it gradually switches to a single peak with the increase of $\bar{\Lambda}^\prime/\bar{\mu}$, which characterizes a strong radiation of the QD. Thus, for the spectrum to be more evident in experiment, a proper designation of the QD orientation and the quadrupole moment is needed.
\begin{figure}[tbp]
\includegraphics[width=\columnwidth]{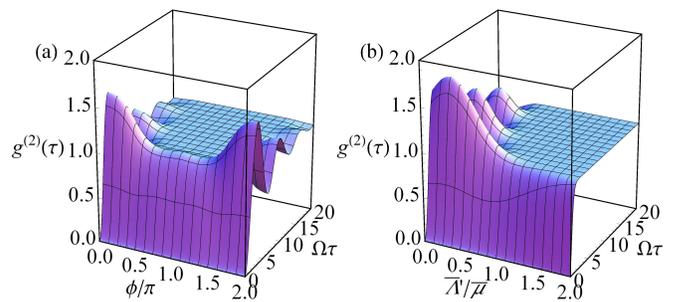}
\caption{$g^{(2)}(\tau )$ versus $\phi$ at $\bar{\Lambda}^\prime/\bar{\mu}=0.5$ (a) and versus $\bar{\Lambda}^\prime/\bar{\mu}$ at $\phi=0$ (b). Parameters are the same as in Fig. \ref{Fig3} and Figs. \ref{Fig4}.} \label{Fig5}
\end{figure}

Another magnitude of the experimental interest is the statistical properties of the emitted light from the QD \cite{Wrigge2008,Ates2009}, which is measured by the second-order correlation function $g^{(2)}(\tau )=\frac{G^{(2)}(\tau )}{\lim_{\tau \rightarrow \infty}G^{(2)}(\tau )}$ with $G^{(2)}(\tau )=\langle \hat{\sigma}_{+}(t)\hat{\sigma}_{+}(t+\tau )\hat{\sigma}_{-}(t+\tau )\hat{\sigma}_{-}(t)\rangle _\text{ss}$. From Eq. (\ref{master}) and with the use of the quantum regression theorem, $g^{(2)}(\tau )$ can be analytically obtained as
\begin{equation}
g^{(2)}(\tau )=1-\Big[\cos (\alpha \tau )+\frac{3\Gamma }{4\alpha }\sin (\alpha \tau)\Big]e^{-3\Gamma \tau /4}.
\end{equation}
One can find that $g^{(2)}(\tau)>g^{(2)}(0)$, which indicates that the probability to detect two emitted photons with time delay $\tau$ is larger than the one without time delay. This is a typical nonclassical property of light, i.e., the antibunching character. It ensures the single-photon nature of the emitted fluorescence. We plot $g^{(2)}(\tau )$ in different QD orientations for $\bar{\Lambda}^\prime/\bar{\mu}=0.5$ in Fig. \ref{Fig5}(a). The $2\pi$ symmetry over the QD orientation is also kept by $g^{(2)}(\tau )$. In addition, $g^{(2)}(\tau )$ experiences from monotonically increase to oscillatory increase in different $\phi$. The oscillation is a manifestation of the laser-driven Rabi oscillation, while the damping of its amplitude is caused by the QD spontaneous emission.  The oscillation of $g^{(2)}(\tau )$ in fixed QD orientation can be dramatically suppressed with the increase of the quadrupole moment [see Fig. \ref{Fig5}(b)]. Such type of transition can be viewed as a manifestation of the fluorescence changed from the weak radiation to the strong radiation regimes due to the presence of the QD mesoscopic effects \cite{Tudela2010}.

Both the spectrum and the second-order correlation function of the fluorescence indicate that one can control the single-photon emission of the QD by its mesoscopic effects, e.g., the spatial orientation and the quadrupole moment.

\section{Conclusions}\label{con}
In summary, we have studied the resonance fluorescence of a mesoscopic QD placed in plasmonic nanostructure. Going beyond the DA, a microscopic description of the decoherence dynamics of the QD has been established. It is revealed that, modified by the QD mesoscopic effects, the spectrum and its statistical property of the resonance fluorescence exhibit different rotation symmetry over the QD orientation and shows significant deviation from those under the DA. The results demonstrate that one can control the interaction between the QD and the surface plasmons by manipulating its mesoscopic effects, which offers a dimension to control the radiation properties of the QD. Our studies are within the present experimental state of the art and instructive for the utilization of the QD mesoscopic characteristics in the nanophotonic device developments.
\section*{Acknowledgments}

This work is supported by the Specialized Research Fund for the Doctoral Program of Higher Education, by the Program for New Century Excellent Talents in University, and by the National Natural Science Foundation of China (Grant No. 11474139).
\appendix
\begin{widetext}
\section{Green's tensor of planar interface}\label{Green}
The Green's tensor in our system is
\begin{equation}
\mathbf{G}(\mathbf{r},\mathbf{r}^{\prime };\omega )=\mathbf{G}_{0}(\mathbf{r},\mathbf{r}^{\prime };\omega )+\mathbf{G}_{\text{R}}(\mathbf{r},\mathbf{r}^{\prime};\omega ),
\end{equation}
where $\mathbf{G}_{0}(\mathbf{r},\mathbf{r}^{\prime };\omega )$ and $\mathbf{G}_{\text{R}}(\mathbf{r},\mathbf{r}^{\prime};\omega )$ denote the contributions of the field propagating in the dielectric and reflected by the metal-dielectric interface, respectively. Under the angular spectrum representation, they take the form \cite{Novotny2006}
\begin{eqnarray}
\mathbf{G}_{0}(\mathbf{r},\mathbf{r}^{\prime };\omega ) &=&\frac{i}{8\pi ^{2}%
} \int_{-\infty }^{\infty }dk_{x}dk_{y}\frac{e^{i[k_{x}(x-x^{\prime
})+k_{y}(y-y^\prime )+k_{z_1}|z-z^\prime|]}}{k_{1}^{2}k_{z_1}}\left(
\begin{array}{ccc}
k_1^{2}-k_{x}^{2} & -k_{x}k_{y} & \mp k_{x}k_{z_1} \\
-k_{x}k_{y} & k_1^{2}-k_{y}^{2} & \mp k_{y}k_{z_1} \\
\mp k_{x}k_{z_1} & \mp k_{y}k_{z_1} & k_1^{2}-k_{z_1}^{2}%
\end{array}%
\right),\label{free-green0}  \\
\mathbf{G}_{\text{R}}(\mathbf{r},\mathbf{r}^{\prime };\omega ) &=&\frac{i}{8\pi ^{2}%
} \int_{-\infty }^{\infty }dk_{x}dk_{y}\frac{e^{i[k_{x}(x-x^\prime
)+k_{y}(y-y^\prime )+k_{z_1}(z+z^\prime )]}}{k_{x}^{2}+k_{y}^{2}} \left(
\mathbf{M}^{\text{s}}+\mathbf{M}^{\text{p}}\right),\label{ref-green0}
\end{eqnarray}
where $\mathbf{k}_1=(k_x,k_y,k_{z_1})$ is the wavevector in the dielectric. The two different signs in $\mathbf{G}_{0}(\mathbf{r},\mathbf{r}^{\prime };\omega )$ are determined by the absolute value of $|z-z^\prime|$, where the upper (lower) sign is applied when $z>z^\prime$ ($z<z^\prime$). In Eq. (\ref{ref-green0}), $\mathbf{G}_{\text{R}}(\mathbf{r},\mathbf{r}^{\prime };\omega ) $ has been splitted into the s-polarized part and p-polarized parts
\begin{equation}
\mathbf{M}^{\text{s}}=\frac{r^{\text{s}}(k_x,k_y)}{k_{z_1}}\left(
\begin{array}{ccc}
k_{y}^{2} & -k_{x}k_{y} & 0 \\
-k_{x}k_{y} & k_{x}^{2} & 0 \\
0 & 0 & 0%
\end{array}%
\right) ,~\mathbf{M}^{\text{p}}=\frac{-r^{\text{p}}(k_x,k_y)}{k_{1}^2}\left(
\begin{array}{ccc}
k_{x}^{2}k_{z_1} & k_{x}k_{y}k_{z_1} & k_{x}(k_{x}^{2}+k_{y}^{2}) \\
k_{x}k_{y}k_{z_1} & k_{y}^{2}k_{z_1} &
k_{y}(k_{x}^{2}+k_{y}^{2}) \\
-k_{x}(k_{x}^{2}+k_{y}^{2}) & -k_{y}(k_{x}^{2}+k_{y}^{2}) &
-(k_{x}^{2}+k_{y}^{2})^{2}/k_{z_1}\end{array}\right).
\end{equation}
where $r^{\text{s}}(k_{x},k_{y})=\frac{\mu_{m}k_{z_{1}}-\mu_{1}k_{z_{\text{m}}}}{\mu_{m}k_{z_{1}}+\mu_{1}k_{z_{\text{m}}}}$ and $r^{\text{p}}(k_{x},k_{y})=\frac{\varepsilon_{\text{m}}k_{z_{1}}-\varepsilon _{1}k_{z_{\text{m}}}}{\varepsilon_{\text{m}}k_{z_{1}}+\varepsilon _{1}k_{z_{\text{m}}}}$ with $k_{z_m}$ being the $z$ component of the wave vector in the metal are the normal Fresnel reflection coefficients for the s-polarized and p-polarized light in the metal-dielectric interface, respectively.
\end{widetext}


\begin{thebibliography}{99}
\bibitem{Haroche2013a} S. Haroche, \href{http://dx.doi.org/10.1103/RevModPhys.85.1083}{Rev. Mod. Phys. \textbf{85}, 1083 (2013)}.

\bibitem{Wineland2013} D. J. Wineland, \href{http://dx.doi.org/10.1103/RevModPhys.85.1103}{Rev. Mod. Phys. \textbf{85}, 1103 (2013)}.

\bibitem{Koelemeij2014} J. C. J. Koelemeij, \href{http://dx.doi.org/10.1038/nphys3113}{Nat. Phys. \textbf{10}, 800 (2014)}.

\bibitem{Bloom2014} B. J. Bloom, T. L. Nicholson, J. R. Williams, S. L. Campbell, M. Bishof, X. Zhang, W. Zhang, S. L. Bromley, and J. Ye, \href{http://dx.doi.org/10.1038/nature12941}{Nature (London) \textbf{506}, 71 (2014)}.


\bibitem{Wang2015} Y. Wang, X.-L. Zhang, T. A. Corcovilos, A. Kumar, and D. S. Weiss, \href{http://dx.doi.org/10.1103/PhysRevLett.115.043003}{Phys. Rev. Lett. \textbf{115}, 043003 (2015)}.

\bibitem{Kato2015} S. Kato, and T. Aoki, \href{http://dx.doi.org/10.1103/PhysRevLett.115.093603}{Phys. Rev. Lett. \textbf{115}, 093603 (2015)}.

\bibitem{Wallraff2004} A. Wallraff, D. I. Schuster, A. Blais, L. Frunzio, R.-S. Huang, J. Majer, S. Kumar, S. M. Girvin, and R. J. Schoelkopf, \href{http://dx.doi.org/10.1038/nature02851}{Nature (London) \textbf{431}, 162 (2004)}.

\bibitem{Guebrou2012} S. Aberra Guebrou, C. Symonds, E. Homeyer, J. C. Plenet, Yu. N. Gartstein, V. M. Agranovich, and J. Bellessa, \href{http://dx.doi.org/10.1103/PhysRevLett.108.066401}{Phys. Rev. Lett. \textbf{108}, 066401 (2012)}.

\bibitem{Niemczyk2010} T. Niemczyk,	F. Deppe, H. Huebl, E. P. Menzel, F. Hocke, M. J. Schwarz, J. J. Garcia-Ripoll, D. Zueco, T. H\"{u}mmer, E. Solano, A. Marx, and R. Gross, \href{http://dx.doi.org/10.1038/nphys1730}{Nat. Phys. \textbf{6}, 772 (2010)}.

\bibitem{Scalari2012} G. Scalari, C. Maissen, D. Tur\v{c}inkov\'{a}, D. Hagenm\"{u}ller, S. De Liberato, C. Ciuti, C. Reichl, D. Schuh, W. Wegscheider, M. Beck, and J. Faist, \href{http://dx.doi.org/10.1126/science.1216022}{Science \textbf{335}, 1323 (2012)}.

\bibitem{Tudela2013} A. Gonz\'{a}lez-Tudela, P. A. Huidobro, L. Mart\'{\i}n-Moreno, C. Tejedor, and F. J. Garc\'{\i}a-Vidal, \href{http://dx.doi.org/10.1103/PhysRevLett.110.126801}{Phys. Rev. Lett. \textbf{110}, 126801 (2013)}.

\bibitem{Cacciola2014}  A. Cacciola, O. D. Stefano, R. Stassi, R. Saija, and S. Savasta, \href{http://dx.doi.org/10.1021/nn504652w}{ACS Nano \textbf{8}, 11483 (2014)}.

\bibitem{Torma2015}  P. T\"{o}rm\"{a} and W. L. Barnes, \href{http://dx.doi.org/10.1088/0034-4885/78/1/013901}{Rep. Prog. Phys. \textbf{78}, 013901 (2015)}.

\bibitem{Andersen2011} M. L. Andersen, S. Stobbe,	A. S. S{\o}rensen and P. Lodahl, \href{http://dx.doi.org/10.1038/nphys1870}{Nat. Phys. \textbf{7}, 215 (2011)}.

\bibitem{Akimov2007} A. V. Akimov, A. Mukherjee, C. L. Yu, D. E. Chang, A. S. Zibrov, P. R. Hemmer, H. Park, and M. D. Lukin, \href{http://dx.doi.org/10.1038/nature06230}{Nature (London) \textbf{450}, 402 (2007)}.

\bibitem{Cronin2009} A. D. Cronin, J. Schmiedmayer, and D. E. Pritchard, \href{http://dx.doi.org/10.1103/RevModPhys.81.1051}{Rev. Mod. Phys. \textbf{81}, 1051 (2013)}.

\bibitem{Haroche2013b} S. Haroche, M. Brune, and J.-M. Rainmond, \href{http://dx.doi.org/10.1063/PT.3.1856}{Phys. Today \textbf{66}, 27 (2013)}.

\bibitem{Muller2007} A. Muller, E. B. Flagg, P. Bianucci, X. Y. Wang, D. G. Deppe, W. Ma, J. Zhang, G. J. Salamo, M. Xiao, and C. K. Shih, \href{http://dx.doi.org/10.1103/PhysRevLett.99.187402}{Phys. Rev. Lett. \textbf{99}, 187402 (2007)}.

\bibitem{Monniello2013} L. Monniello, C. Tonin, R. Hostein, A. Lemaitre, A. Martinez, V. Voliotis, and R. Grousson, \href{http://dx.doi.org/10.1103/PhysRevLett.111.026403}{Phys. Rev. Lett. \textbf{111}, 026403 (2013)}.


\bibitem{Zayatsa2005} A. V. Zayatsa, I. I. Smolyaninovb, A. A. Maradudin, \href{http://dx.doi.org/10.1016/j.physrep.2004.11.001}{Phys. Rep. \textbf{408}, 131 (2005)}.

\bibitem{Yoshie2004} T. Yoshie, A. Scherer, J. Hendrickson, G. Khitrova, H. M. Gibbs, G. Rupper, C. Ell, O. B. Shchekin and D. G. Deppe, \href{http://dx.doi.org/10.1038/nature03119}{Nature (London) \textbf{432}, 200 (2004)}.

\bibitem{Lodahl2015} P. Lodahl, S. Mahmoodian, and S. Stobbe, \href{http://dx.doi.org/10.1103/RevModPhys.87.347}{Rev. Mod. Phys. \textbf{87}, 347 (2015)}.

\bibitem{Ajiki2002} H. Ajiki, T. Tsuji, K. Kawano, and K. Cho, \href{http://dx.doi.org/10.1103/PhysRevB.66.245322}{Phys. Rev. B \textbf{66}, 245322 (2002)}.

\bibitem{Cho2003} K. Cho, \textit{Optical Response of Nanostructures}, (Springer, New York, 2003).

\bibitem{Bamba2008} M. Bamba, and H. Ishihara, \href{http://dx.doi.org/10.1103/PhysRevB.78.085109}{Phys. Rev. B \textbf{78}, 085109 (2008)}.

\bibitem{Iida2009} T. Iida, and H. Ishihara, \href{http://dx.doi.org/10.1002/pssa.200881299}{Phys. Status Solidi A \textbf{206}, 980 (2009)}.

\bibitem{Iida2011} T. Iida, Y. Aiba, and H. Ishihara, \href{http://dx.doi.org/10.1063/1.3551710}{Appl. Phys. Lett. \textbf{98}, 053108 (2011)}.

\bibitem{Takase2013} M. Takase, H. Ajiki, Y. Mizumoto, K. Komeda, M. Nara, H. Nabika, S. Yasuda, H. Ishihara, and K. Murakoshi, \href{http://dx.doi.org/10.1103/PhysRevLett.105.123906}{Nat. Photonics \textbf{7}, 550 (2013)}.

\bibitem{Bamba2010} M. Bamba and H. Ishihara, \href{http://dx.doi.org/10.1103/PhysRevLett.105.123906}{Phys. Rev. Lett. \textbf{105}, 123906 (2010)}.

\bibitem{Osaka2012} Y. Osaka, N. Yokoshi, M. Nakatani, and H. Ishihara, \href{http://dx.doi.org/10.1103/PhysRevLett.112.133601}{Phys. Rev. Lett. \textbf{112}, 133601 (2014)}.


\bibitem{Stobbe2012} S. Stobbe, P. T. Kristensen, J. E. Mortensen, J. M. Hvam, J. M{\o}rk, and P. Lodahl, \href{http://dx.doi.org/10.1103/PhysRevB.86.085304}{Phys. Rev. B \textbf{86}, 085304 (2012)}.

\bibitem{Tighineanu2015} P. Tighineanu, A. S. S{\o}rensen, S. Stobbe, and P. Lodahl, \href{http://dx.doi.org/10.1103/PhysRevLett.114.247401}{Phys. Rev. Lett. \textbf{114}, 247401 (2015)}.

\bibitem{Johansen2008} J. Johansen, S. Stobbe, I. S. Nikolaev, T. Lund-Hansen, P. T. Kristensen, J. M. Hvam, W. L. Vos, and P. Lodahl, \href{http://dx.doi.org/10.1103/PhysRevB.77.073303}{Phys. Rev. B \textbf{77}, 073303 (2008)}.

\bibitem{Leistikow2009} M. D. Leistikow, J. Johansen, A. J. Kettelarij, P. Lodahl, and W. L. Vos, \href{http://dx.doi.org/10.1103/PhysRevB.79.045301}{Phys. Rev. B \textbf{79}, 045301 (2009)}.

\bibitem{Johnson1972} P. B. Johnson, and R.-W. Christy, \href{http://dx.doi.org/10.1103/PhysRevB.6.4370}{Phys. Rev. B \textbf{6}, 4370 (1972)}.

\bibitem{Tudela2010} A. Gonzalez-Tudela, F. J. Rodr\'{\i}guez, L. Quiroga, and C. Tejedor, \href{http://dx.doi.org/10.1103/PhysRevB.82.115334}{Phys. Rev. B \textbf{82}, 115334 (2010)}.

\bibitem{Dung1998} H. T. Dung, L. Kn\"{o}ll, and D.-G. Welsch, \href{http://dx.doi.org/10.1103/PhysRevA.57.3931}{Phys. Rev. A \textbf{57}, 3931 (1998)}.

\bibitem{Sondergaard2001} T. S{\o}ndergaard, and B. Tromborg, \href{http://dx.doi.org/10.1103/PhysRevA.64.033812}{Phys. Rev. A  \textbf{64}, 033812 (2001)}.

\bibitem{Chen2010b} Y. Chen, T. R. Nielsen, N. Gregersen, P. Lodahl, and J. M{\o}rk, \href{http://dx.doi.org/10.1103/PhysRevB.81.125431}{Phys. Rev. B \textbf{81}, 125431 (2010)}.

\bibitem{Cano2010} D. Mart\'{\i}n-Cano, L. Mart\'{\i}n-Moreno, F. J. Garc\'{\i}a-Vidal, and E. Moreno, \href{http://dx.doi.org/10.1021/nl101876f}{Nano Lett. \textbf{10}, 3129 (2010)}.

\bibitem{Dzsotjan2010} D. Dzsotjan, A. S. S{\o}rensen, and M. Fleischhauer, \href{http://dx.doi.org/10.1103/PhysRevB.82.075427}{Phys. Rev. B \textbf{82}, 075427 (2010)}.

\bibitem{Novotny2006} L. Novotny, and B. Hecht,	\textit{Principles of Nano Optics} (Cambridge University Press, Cambridge, UK, 2006).

\bibitem{Zubairy2014} J. Hakami, Ligang Wang, and M. S. Zubairy, \href{http://dx.doi.org/10.1103/PhysRevA.89.053835}{Phys. Rev. A \textbf{89}, 053835 (2014)}.

\bibitem{Pitarke2007} J. M. Pitarke, V. M. Silkin, E. V. Chulkov, and P. M. Echenique, \href{http://dx.doi.org/10.1088/0034-4885/70/1/R01}{Rep. Prog. Phys. \textbf{70}, 1 (2007)}.

\bibitem{Barnes2003} W. L. Barnes, A. Dereux, and T. W. Ebbesen, \href{http://dx.doi.org/10.1038/nature01937}{Nature (London) \textbf{424}, 824 (2003)}.

\bibitem{Garcia2010} F. J. Garcia-Vidal, L. Martin-Moreno, T. W. Ebbesen, and L. Kuipers, \href{http://dx.doi.org/10.1103/RevModPhys.82.729}{Rev. Mod. Phys. \textbf{82}, 729 (2010)}.

\bibitem{Tame2013} M. S. Tame, K. R. McEnery, S. K. \"{O}zdemir, J. Lee, S. A. Maier, and M. S. Kim, \href{http://dx.doi.org/10.1038/nphys2615}{Nat. Phys. \textbf{9}, 329 (2013)}.

\bibitem{Stobbe2009} S. Stobbe, J. Johansen, P. T. Kristensen, J. M. Hvam, and P. Lodahl, \href{http://dx.doi.org/10.1103/PhysRevB.80.155307}{Phys. Rev. B \textbf{80}, 155307 (2009)}.

\bibitem{Sondergaard2004} T. S{\o}ndergaard, and S. I. Bozhevolnyi, \href{http://dx.doi.org/10.1103/PhysRevB.69.045422}{Phys. Rev. B \textbf{69}, 045422 (2004)}.

\bibitem{Siahpoush2012} V. Siahpoush, T. S{\o}ndergaard, and J. Jung, \href{http://dx.doi.org/10.1103/PhysRevB.85.075305}{Phys. Rev. B \textbf{85}, 075305 (2012)}.

\bibitem{Siahpoush2014} V. Siahpoush, and B. Shokri, \href{http://dx.doi.org/10.1016/j.optcom.2013.10.038}{Opt. Commun. \textbf{313}, 315 (2014)}.

\bibitem{Chang2006} D. E. Chang, A. S. S{\o}rensen, P. R. Hemmer, and M. D. Lukin, \href{http://dx.doi.org/10.1103/PhysRevLett.97.053002}{Phys. Rev. Lett. \textbf{97}, 053002 (2006)}.

\bibitem{Tudela2014} A. Gonz\'{a}lez-Tudela, P. A. Huidobro, L. Mart\'{\i}n-Moreno, C. Tejedor, and F. J. Garc\'{\i}a-Vidal, \href{http://dx.doi.org/10.1103/PhysRevB.89.041402}{Phys. Rev. B \textbf{89}, 041402 (2014)}.

\bibitem{Carmichael2000} H. J. Carmichael \textit{Statistical Methods in Quantum Optics 1}, 2nd ed. (Springer, New York, 2000).

\bibitem{Wrigge2008} G. Wrigge, I. Gerhardt, J. Hwang, G. Zumofen, and V. Sandoghdar, \href{http://dx.doi.org/10.1038/nphys812}{Nat. Phys. \textbf{4}, 60 (2008)}.

\bibitem{Ates2009} S. Ates, S. M. Ulrich, S. Reitzenstein, A. L\"{o}ffler, A. Forchel, and P. Michler, \href{http://dx.doi.org/10.1103/PhysRevLett.103.167402}{Phys. Rev. Lett. \textbf{103}, 167402 (2009)}.

\end{thebibliography}
\end{document}